\documentclass[lettersize,journal]{IEEEtran}
\usepackage{amsmath,amsfonts}
\usepackage{algorithmic}
\usepackage{algorithm}
\usepackage{array}
\usepackage[caption=false,font=normalsize,labelfont=sf,textfont=sf]{subfig}
\usepackage{textcomp}
\usepackage{stfloats}
\usepackage{url}
\usepackage{verbatim}
\usepackage{graphicx}
\usepackage{cite}
\usepackage{orcidlink}
\begin{document}

\title{Experimental Demonstration of Computational AoA Detection Using Conformal Frequency Diverse Metasurface Antennas}

\author{Idban Alamzadeh\orcidlink{0000-0003-3493-6117}  ,~\IEEEmembership{Member,~IEEE}, Michael Inman,~\IEEEmembership{Student Member,~IEEE,} and Mohammadreza F. Imani\orcidlink{0000-0003-2619-6358}, ~\IEEEmembership{Member,~IEEE.}
\thanks{The authors are with the School of Electrical, Computer, and Energy Engineering at Arizona State University, Tempe, AZ 85287, USA. (email: Idban.Alamzadeh@asu.edu; Mohammadreza.Imani@asu.edu)}}



\maketitle

\begin{abstract}


Devices that detect angle-of-arrival (AoA) over a wide field of view are crucial for various applications such as wireless communication and navigation. They are often installed on platforms with challenging mechanical and stealth constraints like vehicles, drones, and helmets, where traditional methods—mechanically rotating antennas or conformal arrays—tend to be bulky, heavy, and costly. A recent work has proposed a conformal frequency diverse antenna that is designed to produce angularly diverse patterns that encode angular information into frequency sweeps. This capability allows AoA to be determined across the entire horizon using only two receiving units. This paper experimentally validates this concept, detailing the prototyping process and practical design considerations. The AoA detection capabilities of the proposed device are confirmed through experimental demonstrations. The proposed conformal metasurfaces offer an alternative hardware solution for sensing over large fields of view, with potential applications in radar sensing, situational awareness, and navigation.



\end{abstract}


\section{Introduction}
\IEEEPARstart{D}{evices}  that can detect AoA of incident signals in a large field of view with high resolution can greatly benefit various wireless communication and sensing systems \cite{adams2012robotic,klausing1989feasibility,vivet2013localization,ali2014rotating,wang2022airport,lee2021dual,angelilli2017family,peshwe2022threat,meyer2019automotive,venon2022millimeter,grebner2022radar,rouveure2018description,nan2022panoramic,stockel2022high}. These devices can be implemented using a mechanically rotating antenna \cite{klausing1989feasibility,vivet2013localization,ali2014rotating,wang2022airport}, which can provide excellent performance, as they establish a one-to-one relationship between each measurement and the corresponding angle or pixel. However, rotating antenna systems tend to be bulky and heavy, making it difficult to conform to arbitrary shapes with stealth and mechanical restrictions. An alternative solution involves using conformal antenna arrays that can be electronically switched on and off or fed through specialized feeds \cite{angelilli2017family,mohammadi2017direction,jackson20222d}. This method performs well in interrogating a large field of view (e.g. the entire azimuth plane). However, it is often costly and complicated due to the large number of transceivers or the intricate feed structures involved. Moreover, achieving optimal performance necessitates careful management of trade-offs between element gain and array spacing \cite{jackson2014direction,nechaev2017evaluation}. To mitigate these challenges, compressive sensing algorithms or machine learning techniques have been proposed to reduce the number of measurements needed, particularly with pattern reconfigurable antennas \cite{shen2016underdetermined,uemura2021direction,schab2013direction,friedrichs2021compact}. Yet, these approaches still rely on conventional antenna arrays, which come with the previously mentioned issues.

In this paper, we experimentally demonstrate a conformal multiplexing metasurface antenna to simplify the hardware for detecting the AoA. The idea of using multiplexing metasurfaces for computational sensing has gained traction since the work in \cite{hunt2013metamaterial}. In this work, a planar metamaterial aperture was designed to generate frequency-diverse illumination patterns. These patterns encoded the spatial information of an imaging domain into rapid frequency sweeps. By applying compressive reconstruction techniques to the collected data, reflectivity maps of the scene were retrieved. These works differed from traditional planar microwave imaging systems: no one-to-one relationship existed between the acquired data and the probed point. In contrast, the frequency-diverse patterns were spatially diverse and illuminated the whole region of interest. This groundbreaking work paved the way for a simple flat security screening system capable of imaging human-sized objects at video-frame rate without requiring any mechanical scanning \cite{Ref130}. Multiplexing metasurface antennas thus promises to expand the design space for microwave sensing \cite{imani2020review}. Planar multiplexing metasurfaces have also been used for simplifying AoA detection devices\cite{li2023direct,hoang2021single,abbasi2021lens,lin2021single,yurduseven2019frequency,li2024frequency}.

Extending this transforming approach to conformal configurations can thus address many of the challenges of conventional systems. In \cite{imani2023conformal}, a conformal frequency--diverse metasurface antenna was proposed, which was distinct from previous conformal antennas, e.g. impedance-modulated and holographic surfaces \cite{gregoire20133,wang2023multi,ramalingam2017axially,longhi2023array}, composite right-hand left-hand metasurfaces \cite{hashemi2009electronically}, and leaky wave antennas \cite{gomez2011analysis,lee20232}. In all those works, the main objective is to generate directive beams toward a given direction. In \cite{imani2023conformal}, the conformal metasurface was designed to generate angularly diverse patterns that change as a function of frequency. The frequency-diverse patterns encoded the information about the incident signal into simple frequency sweeps at two single ports. Using full-wave simulations, it was shown that the data collected by this device can be computationally processed to detect the AoA of signals over the whole horizon.  

In this letter, we experimentally demonstrate this concept. We detail the design considerations for a frequency-diverse conformal metasurface antenna and explain the trade-offs governing the design space. A proof-of-concept prototype will be fabricated. Our experiments show that the proposed design can generate angularly diverse patterns that vary with frequency. We also outline a simple computational process for analyzing the data collected by the conformal metasurface to detect AoAs across the entire horizon. Lastly, we experimentally validate the ability to detect AoA for sources at varying distances. 
\section{Frequency-diverse conformal metasurface}
\begin{figure}[!t]
\centering
\includegraphics[width=0.75\linewidth]{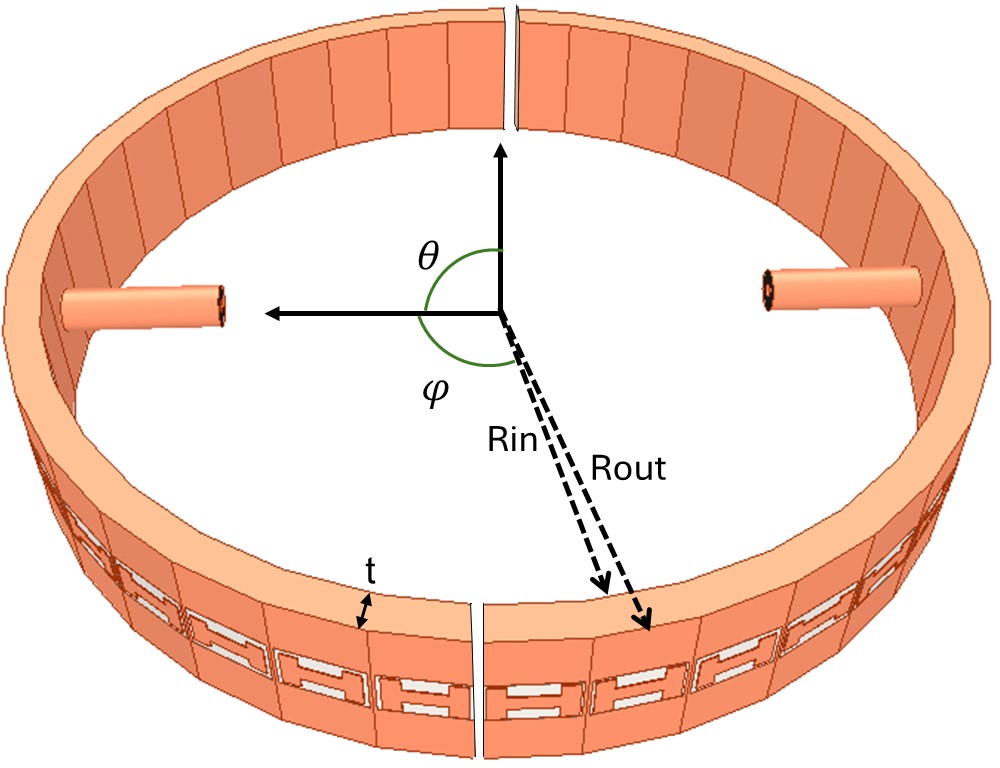}

\caption{\label{cfdma5} The conformal frequency-diverse metasurface antennas.}
\end{figure}

The general configuration of the proposed conformal frequency-diverse metasurface antenna is shown in Fig. \ref{cfdma5}. It comprises two conformal substrate integrated waveguides (SIW) that wrap around a cylindrical structure. In this figure and the experimental setup, we use a cylinder with a circular cross-section, but the proposed operation can be extended to arbitrary shapes. Each SIW covers half of the cylinder and is connected to a coaxial connector. The top conducting walls of the SIWs are fashioned with metamaterial radiators whose resonance frequencies are selected randomly within a band of operation (X-band here). The overall radiation/reception pattern of this structure is the superposition of the field generated/received by the metamaterial elements. Given the random distribution of elements' resonance frequencies, this structure generates angularly diverse patterns that change with frequency. The AoA of the incident signal can thus be deduced from the frequency measurement at the coaxial connectors. It is worth noting that the number of antennas and the cylinder cross-section can be different in other practical implementations. The choices here are to simplify the proof of concept experiment.

\begin{figure}
\centering
\includegraphics[width=0.9\linewidth]{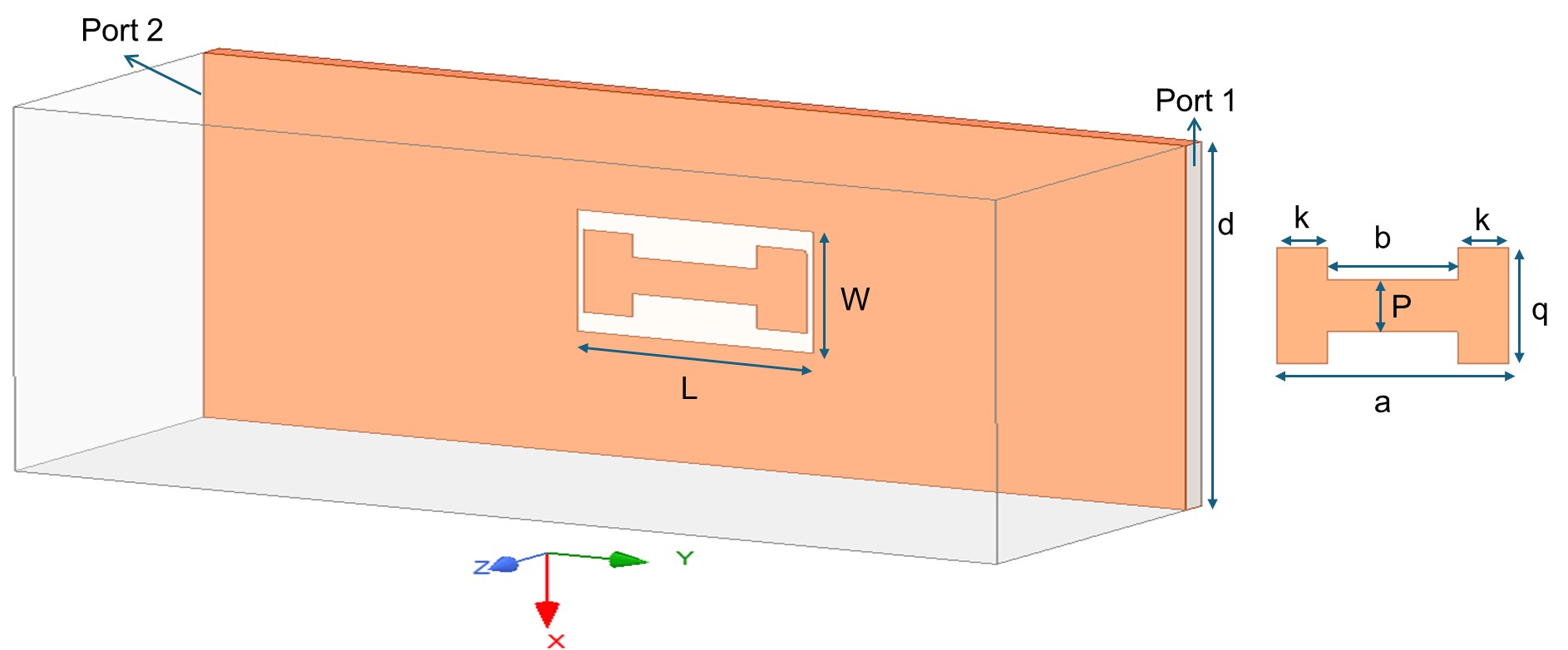}

\caption{\label{cfdma1} The simulation setup for designing the metamaterial elements compromising the frequency-diverse metasurface antennas shown in Fig. \ref{cfdma5}. The SIW is terminated in waveguide ports 1 and 2.}
\end{figure}

To design this structure, we start with the constitutive elements. In \cite{imani2023conformal}, I-shaped slots with different resonance frequencies were used. Due to the orientation of these elements, which had a large opening perpendicular to the direction of the guided waves within the SIW, they tended to cause significant perturbation to the guided mode. As a result, they usually exhibited high reflection, even away from resonances. Near resonance, these elements often radiated considerable power, leading to the depletion of the guided wave. Here, we use complementary elements with I-shaped conductors placed within a rectangular slot, as shown in Fig. \ref{cfdma1}. We can limit the transverse opening in this geometry to reduce the guided mode’s perturbation. We analyze these elements in a planar configuration \cite{sleasman2015waveguide} to design them. In this configuration, we model the sidewalls of the SIW with perfect conducting boundary conditions for simplicity. We assume the structure is implemented on a $30$-mil Rogers Duroid $5880$ substrate. This substrate is flexible and can be bent into different shapes. The width of this SIW is $d = 15$ mm, corresponding to a guided mode cutoff frequency of $6.74$ GHz.

The dimensional details of the elements are highlighted in Fig.~\ref{cfdma1}. By running parametric sweeps of the meta-atom’s geometrical parameters, we are able to design elements with different resonance frequencies empirically. The main design criteria are as follows: the elements should exhibit resonance in the X band while maintaining low reflection ($S_{11}$) and coupling ($S_{21}$). A few geometrical parameters, namely $q = 2.6$ mm and $k = 1.5$ mm were kept fixed. By fixing the element's width, $q$, we limit the perturbation of the guided mode (or, equivalently, reduce reflection off-resonance) and eliminate energy depletion. To simplify fabrication, we also kept the narrowest gap at $\beta = 0.2$ mm. The total size of the slot containing the I-shape meta-atom is then $L = a + 2 \beta$ and $W = q + 2\beta$. During the design process, the dimensions $a$ and $p$ are swept, within the ranges $a = [6, 7]$ mm and $p = [0.5,1]$ mm, to find different resonance frequencies. It is worth noting that $b$ changes as we change $a$.  




\begin{figure}
\centering
\includegraphics[width=\linewidth]{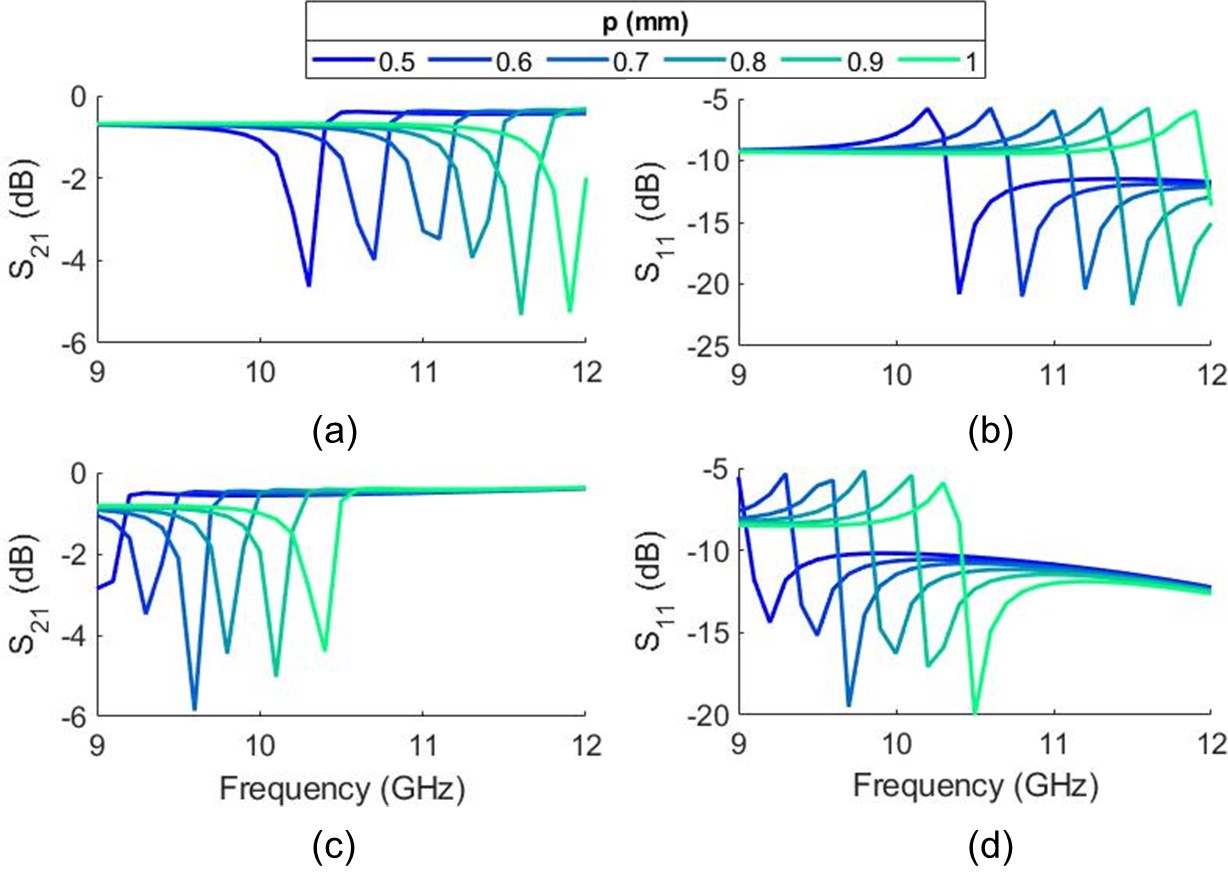}
\caption{\label{cfdma2} Left: $S_{21}$ and right: $S_{11}$ plots of the elements for the set of $p$ values when the value of $a$ is a) - b) $6$ mm and c) - d) $7$ mm.}
\end{figure}

 The S-parameters of the meta-atom while varying $a$ and $p$ over the defined range are shown in Fig.~\ref{cfdma2}. For illustrations, we selected two specific values of $a$ and presented S-parameter analyses as we changed $p$. In this Figure, a dip in $S_{21}$ denotes a resonance. We want this dip not to be too low (i.e., high reflection/radiation) or too high (i.e., very low coupling); in other words, striking a balance between inter-element wave propagation and the radiation level at each element. For this proof of concept design, we consider a value above $-6$ dB to be acceptable, as shown in Fig.~\ref{cfdma2}. In this figure, we have also analyzed the reflection parameter, $S_{11}$. Our goal here is to ensure the element does not exhibit high reflection, both at the resonance and far away from it. In this case, as long as the reflection coefficient stayed below $-5$ dB, we considered it to meet our design expectations. The above levels of $S_{21}$ and $S_{11}$ were selected based on a rough calculation to ensure the guided wave gets depleted, considering the small number of elements used in the proof of concept demonstration presented in the next section. In practice, they should be tailored in a case-specific manner to fulfill the design criteria of respective applications, and the values selected here are for the proof-of-concept and should not be considered optimum. For example, a design with more elements should use a larger $S_{21}$ to ensure the wave reaches all elements (e.g. by placing them off center).
 
 Both $S_{11}$ and $S_{21}$ parameters exhibit shifts in the resonance frequency of the frequency-diverse meta-atom. This behavior is critical for ensuring the frequency-diverse responses of the meta-atoms constituting the conformal metasurface. The radiated power at the resonance is around $30-40$\%.



\section{Experimental Results}
As can be seen in Figs. \ref{cfdma2}, we can cover the whole band by changing only two geometrical parameters. Using these simulations as the guide, we select meta-atoms with varying resonance frequencies and distribute them randomly around the conformal metasurface as depicted in Fig.~\ref{cfdma5}. In doing so, we ensure the elements along each quarter cover the whole band of operation. While the distribution of elements can be potentially optimized (an interesting direction for future work), our approach is to use a random distribution due to its simplicity. Bending the SIW will slightly change the resonant frequencies of the elements. However, given the elements' subwavelength size and the large curvature radius, this impact can be expected to be around $1$\%, as shown in \cite{bayraktar2012slots}. We use two semi-circular conformal metasurfaces to cover the whole circumference of the cylindrical structure. The outer radius of curvature is $R_{out} = 45$ mm. Each metasurface antenna consists of $17$ metamaterial elements excited near the center using a coaxial connector. We alter the distribution of elements in each half on either side of the feed for each metasurface. These design choices were made to increase pattern diversity. The optimum selection of elements and their distribution is a subject for future work. To fabricate the metasurface, we sent our design blueprint to a professional PCB manufacturer. Once the PCB of the prototype was ready, we soldered an off-the-shelf coaxial connector to each semi-circular metasurface. Each metasurface is terminated in a conducting wall at either end. For this proof-of-concept demonstration, we did not use via walls to reduce cost and complexity. Instead, we covered the metasurfaces' walls using copper tape. 

\begin{figure}[!t]
\centering
\includegraphics[width=0.675\linewidth]{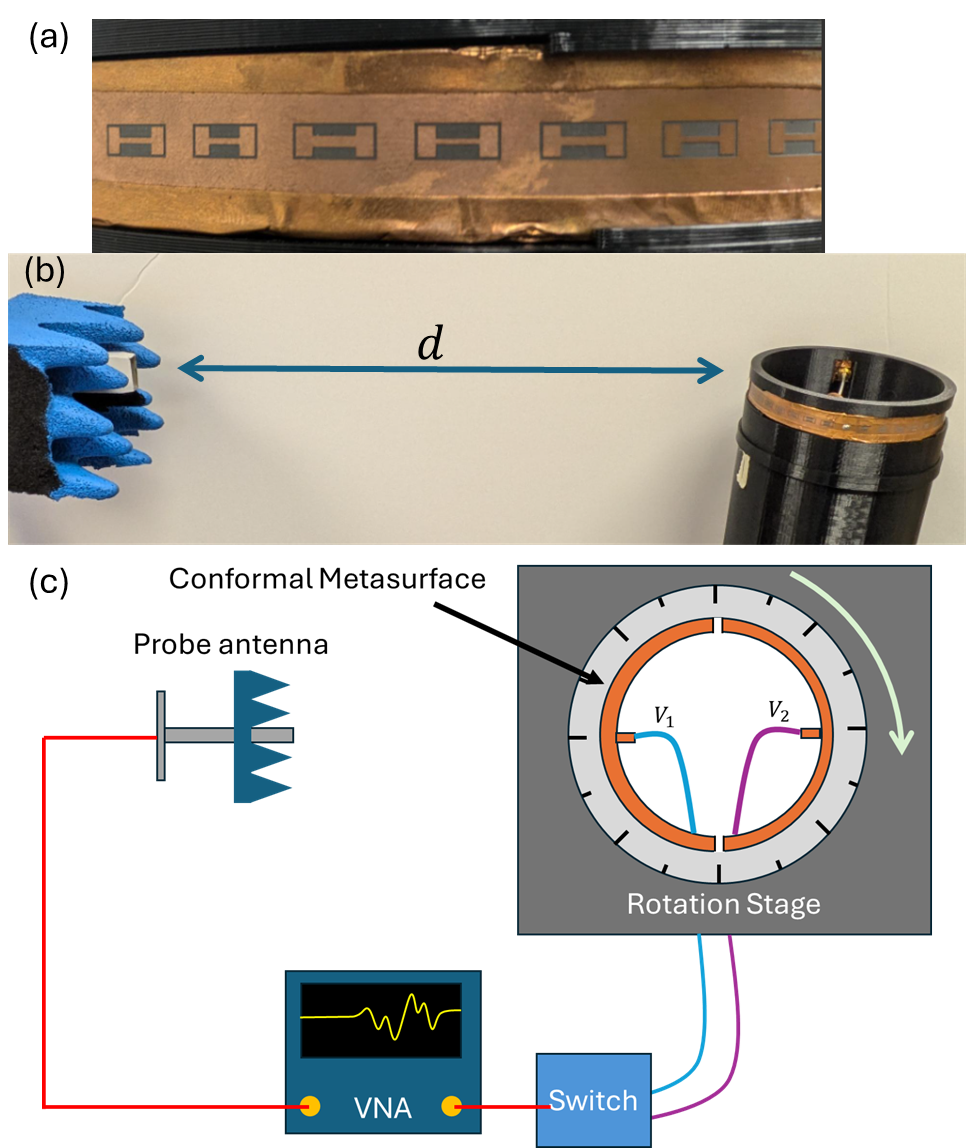}
\caption{\label{cfdma_exp} (a) Close-up view of the fabricated conformal metasurface. (b) Experimental setup. (c) Experiment configuration schematic.}
\vspace{-2mm}
\end{figure}

\begin{figure}[!t]
\centering
\includegraphics[width=0.9\linewidth]{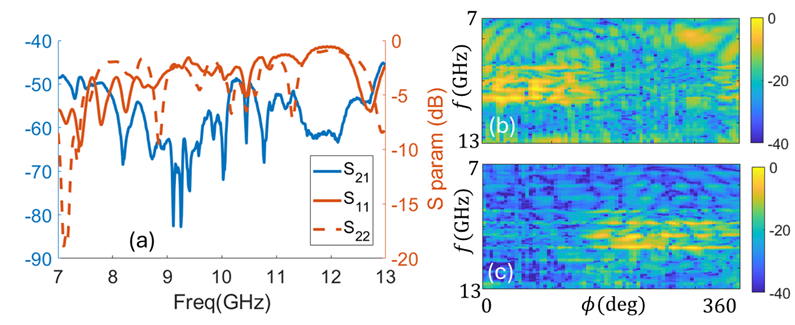}
\caption{\label{cfdma_exp_Sparam} Measured (a) scattering parameters, (b) and (c) radiation/reception patterns of the fabricated conformal metasurfaces.}
\vspace{-3mm}
\end{figure}

The fabricated sample, the experimental setup, and a close-up of a few fabricated meta-atoms are shown in Fig. \ref{cfdma_exp}. The variation in the elements' geometrical properties is evident in Fig.~\ref{cfdma_exp}a. To experimentally verify this structure, we place it on a rotation stage using a cylindrical 3D-printed mount that can hold the metasurface in place (see Figs. \ref{cfdma_exp} b and c). The rotation stage is used here to accurately measure the direction of a transmitter (ground truth). The proposed system can operate without any moving parts in practice. First, we measure the metasurfaces' scattering parameters when bent to conform to the cylindrical mount. For this measurement, we directly connect the VNA ports to the antennas (without the mechanical switch). The measured S parameters are plotted in Fig. \ref{cfdma_exp_Sparam}. We see both the reflection coefficient and the isolation/coupling exhibit dips, especially in the $8.5-11.5$ GHz range, confirming the presence of resonances. The overall band of operation seems to be slightly shifted, which can be attributed to fabrication tolerances and coupling between elements. The reflection coefficient is above -10 dB, which is typically desired, due to the imperfect transition from the coaxial connector to the SIW. While this is acceptable for the proof-of-concept, future designs could utilize more advanced configurations, like aperture coupling. As shown in Fig. \ref{cfdma_exp_Sparam}, the direct coupling between the two metasurfaces is, on average, less than $-40$ dB.

Each metasurface is then connected (via a rotary joint) to one input port of a mechanical switch. The output port of the mechanical switch is connected to port $1$ of a vector network analyzer (VNA). The port $2$ of the VNA is connected to an X-band standard open-ended waveguide probe. Using this setup, we measured each antenna's radiation/reception patterns at $d_{ref}=16$ cm, as shown in Fig. \ref{cfdma_exp_Sparam}. We see that this antenna's patterns change drastically as a function of frequency. It is also evident that the antennas generate angularly selective patterns. Examining this figure, we also conclude that the two antennas radiate with higher energy within the $8.5-11.5$ GHz band.

To use the diverse frequency patterns of the conformal metasurfaces for AoA detection, we use a similar procedure as in \cite{imani2023conformal}. In this framework, we discretize the possible AoA over the whole horizon into $N$ bins. The center of the bins would be reference angles. We populate a sensing matrix by measuring signal at the ports of the conformal metasurfaces for those $N$ reference angles over $M$ frequency points within a band of operation. For this purpose, let us refer to the measurement for one antenna as $V_1$ and the other as $V_2$. The sensing matrix, $\mathbf{H}_{M\times N}$, is then defined using cross-correlation of those measurements:

\begin{equation}
\label{data_g}
H(i,j) = V_1 V_2^*.
\end{equation}

Where $*$ is the complex conjugate. Here $i$  is the frequency index and $j$ is the angle index. By using cross-correlation, we ensure the reference matrix is independent of the measurements phase reference (or the distance, $d$). In our proof-of-concept demonstration, we use steps of $5^\circ$, resulting in $N=72$. This sensing matrix is populated using the setup shown in Fig. \ref{cfdma_exp}. We use a distance $d_{ref}=16$ cm for reference measurements. Next, we conduct the same measurement at different distances. We will denote those measurements with $\mathbf{g}_{M\times1}$. We can relate these measurements to the AoA using:
\begin{equation}
\label{ghf}
\mathbf{g} = \mathbf{Hf}.
\end{equation}
where $\mathbf{f}_{N\times 1}$ is a matrix whose $i$th entry is $1$ if the incident signal is from the $i$th reference angle and $0$ otherwise. Given the measured signal, $\mathbf{g}$ and precharacterized sensing matrix $\mathbf{H}$, we can solve (\ref{ghf}) to find an estimate of $\mathbf{f}$, which we will denote as $\mathbf{f}_{est}$. Since $\mathbf{H}$ is not a square matrix, we solve (\ref{ghf}) using computational techniques. Here, we used MATLAB's built-in solver, \textit{cgs}. However, the resulting $\mathbf{f}_{est}$ will not be 0 and 1. Instead, we consider the estimated AoA to be the reference angle at which $|\mathbf{f}_{est}|$ exhibits a maximum.

\begin{figure}[!t]
\centering
\includegraphics[width=0.85\linewidth]{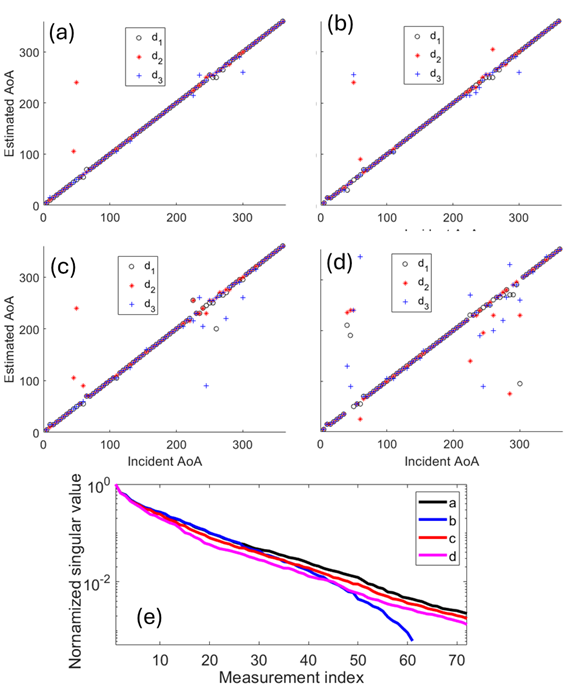}
\caption{\label{cfdma_exp_AoA} Estimated versus incident AoA for $d_1=21$ cm, $d_2=26$ cm, and $d_3=33.5$ cm. (a) Using $M=301$ frequency points over $8.5-11.5$ GHz. (b) Using $M=61$ frequency points over $8.5-11.5$ GHz. (c) Using $M=201$ frequency points over $9-11$ GHz. (d) Using $M=151$ frequency points over $9.25-10.75$ GHz. (e) Corresponding normalized SVD of $\mathbf{H}$.}
\vspace{-3mm}
\end{figure}

The results of the AoA estimation for measurements taken at various distances and settings are displayed in Fig. \ref{cfdma_exp_AoA}. In Fig. \ref{cfdma_exp_AoA}a, we present the case where the entire useful bandwidth of $8.5-11.5$ GHz is utilized, using \(M = 301\) frequency points. It is evident that the AoA can be detected at all distances. However, some errors become apparent near the angle where the incident signal interacts primarily with one of the conformal metasurfaces. Consequently, the resulting cross-correlation is weak and more vulnerable to noise. We also observe a similar trend as the distance increases, where the error may rise due to a reduced signal-to-noise ratio (SNR). 

It is important to note that some seemingly erroneous detections may stem from slight misalignments that occur as the distance increases, resulting in a few degrees of error. In Fig. \ref{cfdma_exp_AoA}b, we reduce the number of frequency points to \(M = 61\) while maintaining the same bandwidth. It is evident that the performance slightly degrades, but the results remain satisfactory. Further reduction of frequency points may increase the difficulty of the inverse problem and could necessitate the use of alternative compressive detection techniques. In Figs. \ref{cfdma_exp_AoA} c and d, we reduced the bandwidth, showing a slight performance degradation as the bandwidth decreases. However, decreasing the bandwidth below $1.5$ GHz can significantly impact performance. The trend we observe in these results is also consistent with the changes in the singular value decomposition (SVD) of the sensing matrix $\mathbf{H}$, shown in Fig. \ref{cfdma_exp_AoA}e. For narrowband cases, we need to redesign the conformal metasurface with a much higher quality factor or utilize disordered SIWs. Alternatively, we can use reconfigurable conformal metasurface antennas \cite{alamzadeh2024computational}. It is worth noting that the experimental trends observed in Fig. \ref{cfdma_exp_AoA} align perfectly with the theoretical predictions in \cite{imani2023conformal}.



\section{Conclusion}
We presented experimental verification of computational AoA detection using conformal frequency-diverse metasurface antennas. By conducting a cylindrical scan, we demonstrated that these antennas generate diverse radiation patterns that change as a function of frequency. We outlined a processing technique to utilize these frequency-diverse, angularly selective patterns for AoA detection. Our experimental setup further confirmed the capability to detect AoA at varying distances. Additionally, we investigated the impact of different frequency points and bandwidths on detection performance. Conformal frequency-diverse metasurfaces have potential applications in wireless sensing, direction finding, and navigation.

\section*{Acknowledgments}
This material is based upon work supported by the National Science Foundation under Grant No. ECCS-2333023.


{\small

\bibliographystyle{IEEEtran}
\bibliography{reference.bib}

\begin{thebibliography}{10}
\providecommand{\url}[1]{#1}
\csname url@samestyle\endcsname
\providecommand{\newblock}{\relax}
\providecommand{\bibinfo}[2]{#2}
\providecommand{\BIBentrySTDinterwordspacing}{\spaceskip=0pt\relax}
\providecommand{\BIBentryALTinterwordstretchfactor}{4}
\providecommand{\BIBentryALTinterwordspacing}{\spaceskip=\fontdimen2\font plus
\BIBentryALTinterwordstretchfactor\fontdimen3\font minus \fontdimen4\font\relax}
\providecommand{\BIBforeignlanguage}[2]{{%
\expandafter\ifx\csname l@#1\endcsname\relax
\typeout{** WARNING: IEEEtran.bst: No hyphenation pattern has been}%
\typeout{** loaded for the language `#1'. Using the pattern for}%
\typeout{** the default language instead.}%
\else
\language=\csname l@#1\endcsname
\fi
#2}}
\providecommand{\BIBdecl}{\relax}
\BIBdecl

\bibitem{adams2012robotic}
M.~Adams and M.~D. Adams, \emph{Robotic navigation and mapping with radar}.\hskip 1em plus 0.5em minus 0.4em\relax Artech House, 2012.

\bibitem{klausing1989feasibility}
H.~Klausing, ``Feasibility of a synthetic aperture radar with rotating antennas (rosar),'' in \emph{1989 19th European Microwave Conference}.\hskip 1em plus 0.5em minus 0.4em\relax IEEE, 1989.

\bibitem{vivet2013localization}
D.~Vivet, P.~Checchin, and R.~Chapuis, ``Localization and mapping using only a rotating fmcw radar sensor,'' \emph{Sensors}, vol.~13, no.~4, pp. 4527--4552, 2013.

\bibitem{ali2014rotating}
F.~Ali, G.~Bauer, and M.~Vossiek, ``A rotating synthetic aperture radar imaging concept for robot navigation,'' \emph{IEEE Trans. on microwave theory and techniques}, vol.~62, no.~7, pp. 1545--1553, 2014.

\bibitem{wang2022airport}
Y.~Wang, Q.~Song, J.~Wang, and H.~Yu, ``Airport runway foreign object debris detection system based on arc-scanning sar technology,'' \emph{IEEE Trans. on Geoscience and Remote Sensing}, vol.~60, pp. 1--16, 2022.

\bibitem{lee2021dual}
S.~Lee, S.-Y. Kwon, B.-J. Kim, H.-S. Lim, and J.-E. Lee, ``Dual-mode radar sensor for indoor environment mapping,'' \emph{Sensors}, vol.~21, no.~7, p. 2469, 2021.

\bibitem{angelilli2017family}
M.~Angelilli, L.~Infante, and P.~Pacifici, ``A family of secondary surveillance radars based on conformal antenna array geometries,'' in \emph{2017 IEEE Radar Conference}.\hskip 1em plus 0.5em minus 0.4em\relax IEEE, 2017, pp. 1681--1684.

\bibitem{peshwe2022threat}
P.~D. Peshwe, A.~G. Kothari, I.~S. Darwhekar, and A.~M. Chauhan, ``Threat detection with millimeter wave conformal antenna array using beamforming and direction of arrival estimation,'' \emph{International Journal of RF and Microwave Computer-Aided Engineering}, vol.~32, no.~3, p. e23030, 2022.

\bibitem{meyer2019automotive}
M.~Meyer and G.~Kuschk, ``Automotive radar dataset for deep learning based 3d object detection,'' in \emph{2019 16th european radar conference (EuRAD)}.\hskip 1em plus 0.5em minus 0.4em\relax IEEE, 2019, pp. 129--132.

\bibitem{venon2022millimeter}
A.~Venon, Y.~Dupuis, P.~Vasseur, and P.~Merriaux, ``Millimeter wave fmcw radars for perception, recognition and localization in automotive applications: A survey,'' \emph{IEEE Trans. on Intelligent Vehicles}, vol.~7, no.~3, pp. 533--555, 2022.

\bibitem{grebner2022radar}
T.~Grebner, P.~Schoeder, V.~Janoudi, and C.~Waldschmidt, ``Radar-based mapping of the environment: Occupancy grid-map versus sar,'' \emph{IEEE Microw. Wirel. Compon. Lett.}, vol.~32, no.~3, pp. 253--256, 2022.

\bibitem{rouveure2018description}
R.~Rouveure, P.~Faure, and M.-O. Monod, ``Description and experimental results of a panoramic k-band radar dedicated to perception in mobile robotics applications,'' \emph{Journal of Field Robotics}, vol.~35, no.~5, pp. 678--704, 2018.

\bibitem{nan2022panoramic}
Y.~Nan, X.~Huang, and Y.~J. Guo, ``A panoramic synthetic aperture radar,'' \emph{IEEE Trans. on Geoscience and Remote Sensing}, vol.~60, pp. 1--13, 2022.

\bibitem{stockel2022high}
P.~Stockel, P.~Wallrath, N.~Pohl, and R.~Herschel, ``High accuracy position calculation of a hovering uav using a rotating radar,'' in \emph{2022 19th European Radar Conference (EuRAD)}.\hskip 1em plus 0.5em minus 0.4em\relax IEEE, 2022, pp. 129--132.

\bibitem{mohammadi2017direction}
S.~Mohammadi, A.~Ghani, and S.~H. Sedighy, ``Direction-of-arrival estimation in conformal microstrip patch array antenna,'' \emph{IEEE Trans. on Antennas Propag.}, vol.~66, no.~1, pp. 511--515, 2017.

\bibitem{jackson20222d}
B.~R. Jackson, ``2d direction of arrival estimation using uniform circular arrays with radiation pattern reconfigurable antennas,'' \emph{IEEE Access}, vol.~10, pp. 11\,909--11\,923, 2022.

\bibitem{jackson2014direction}
B.~R. Jackson, S.~Rajan, B.~J. Liao, and S.~Wang, ``Direction of arrival estimation using directive antennas in uniform circular arrays,'' \emph{IEEE Trans. on Antennas and Propag.}, vol.~63, no.~2, pp. 736--747, 2014.

\bibitem{nechaev2017evaluation}
Y.~B. Nechaev and I.~Peshkov, ``Evaluation of the influence of directivity factor of directive elements of conformal antenna arrays on the performances of azimuth-elevation doa estimation,'' in \emph{2017 PIERS Symposium-Spring}.\hskip 1em plus 0.5em minus 0.4em\relax IEEE, 2017, pp. 490--495.

\bibitem{shen2016underdetermined}
Q.~Shen, W.~Liu, W.~Cui, and S.~Wu, ``Underdetermined doa estimation under the compressive sensing framework: A review,'' \emph{IEEE Access}, vol.~4, pp. 8865--8878, 2016.

\bibitem{uemura2021direction}
S.~Uemura, K.~Nishimori, R.~Taniguchi, M.~Inomata, K.~Kitao, T.~Imai, S.~Suyama, H.~Ishikawa, and Y.~Oda, ``Direction-of-arrival estimation with circular array using compressed sensing in 20 ghz band,'' \emph{IEEE Antennas Wirel. Propag. Lett}, vol.~20, no.~5, pp. 703--707, 2021.

\bibitem{schab2013direction}
K.~Schab, E.~L. Daly, and J.~T. Bernhard, ``Direction estimation using compressive array sensing and pattern reconfigurable antennas,'' in \emph{Asilomar Conference on Signals, Systems and Computers}.\hskip 1em plus 0.5em minus 0.4em\relax IEEE, 2013, pp. 927--930.

\bibitem{friedrichs2021compact}
G.~R. Friedrichs, M.~A. Elmansouri, and D.~S. Filipovic, ``A compact machine learning architecture for wideband amplitude-only direction finding,'' \emph{IEEE Trans. on Antennas Propag.}, vol.~70, no.~7, pp. 5189--5198, 2021.

\bibitem{hunt2013metamaterial}
J.~Hunt, T.~Driscoll, A.~Mrozack, G.~Lipworth, M.~Reynolds, D.~J. Brady, and D.~R. Smith, ``Metamaterial apertures for computational imaging,'' \emph{Science}, vol. 339, no. 6117, pp. 310--313, 2013.

\bibitem{Ref130}
J.~Gollub, O.~Yurduseven, K.~Trofatter, D.~Arnitz, M.~Imani \emph{et~al.}, ``Large metasurface aperture for millimeter wave computational imaging at the human-scale,'' \emph{Sci. Rep.}, vol.~7, p. 42650, 2017.

\bibitem{imani2020review}
M.~F. Imani, J.~N. Gollub, O.~Yurduseven, A.~V. Diebold, M.~Boyarsky, T.~Fromenteze, L.~Pulido-Mancera, T.~Sleasman, and D.~R. Smith, ``Review of metasurface antennas for computational microwave imaging,'' \emph{IEEE Trans. on Antennas Propag.}, vol.~68, no.~3, pp. 1860--1875, 2020.

\bibitem{li2023direct}
W.~Li, N.~Wang, and J.~Qi, ``Direct angle of arrival (aoa) estimation using a metasurface antenna with single frequency phaseless measurements obeyed schwarz inequality,'' \emph{IEEE Trans. on Microwave Theory and Techniques}, vol.~72, no.~4, pp. 2677--2685, 2023.

\bibitem{hoang2021single}
T.~V. Hoang, V.~Fusco, M.~A.~B. Abbasi, and O.~Yurduseven, ``Single-pixel polarimetric direction of arrival estimation using programmable coding metasurface aperture,'' \emph{Sci. Rep.}, vol.~11, no.~1, p. 23830, 2021.

\bibitem{abbasi2021lens}
M.~A.~B. Abbasi, V.~Fusco, O.~Yurduseven, R.~I. Ansari, and S.~L. Cotton, ``Lens-loaded cavity antenna with detector diode as a direction-of-arrival estimator,'' \emph{IEEE Antennas Wirel. Propag. Lett}, vol.~20, no.~11, pp. 2176--2180, 2021.

\bibitem{lin2021single}
M.~Lin, M.~Xu, X.~Wan, H.~Liu, Z.~Wu, J.~Liu, B.~Deng, D.~Guan, and S.~Zha, ``Single sensor to estimate doa with programmable metasurface,'' \emph{IEEE Internet of Things Journal}, vol.~8, no.~12, pp. 10\,187--10\,197, 2021.

\bibitem{yurduseven2019frequency}
O.~Yurduseven, M.~A.~B. Abbasi, T.~Fromenteze, and V.~Fusco, ``Frequency-diverse computational direction of arrival estimation technique,'' \emph{Sci. Rep.}, vol.~9, no.~1, p. 16704, 2019.

\bibitem{li2024frequency}
M.~S. Li, M.~Abdullah, J.~He, K.~Wang, C.~Fumeaux, and W.~Withayachumnankul, ``Frequency-diverse antenna with convolutional neural networks for direction-of-arrival estimation in terahertz communications,'' \emph{IEEE Trans. on Terahertz Science and Technology}, vol.~14, no.~3, pp. 354--363, 2024.

\bibitem{imani2023conformal}
M.~F. Imani and I.~Alamzadeh, ``Conformal frequency-diverse metasurface for computational {AoA} detection,'' \emph{IEEE Antennas Wirel. Propag. Lett}, vol.~22, no.~11, pp. 2634--2638, 2023.

\bibitem{gregoire20133}
D.~J. Gregoire, ``3-d conformal metasurfaces,'' \emph{IEEE Antennas Wirel. Propag. Lett}, vol.~12, pp. 233--236, 2013.

\bibitem{wang2023multi}
Y.~Wang, Q.~Feng, X.~Kong, H.~Liu, J.~Han, and L.~Li, ``Multi-feed beam-switchable cylindrical conformal holographic metasurface antenna,'' \emph{IEEE Antennas Wirel. Propag. Lett}, 2023.

\bibitem{ramalingam2017axially}
S.~Ramalingam, C.~Balanis, C.~R. Birtcher, S.~Pandi, and H.~N. Shaman, ``Axially modulated cylindrical metasurface leaky-wave antennas,'' \emph{IEEE Antennas Wirel. Propag. Lett}, vol.~17, no.~1, pp. 130--133, 2017.

\bibitem{longhi2023array}
M.~Longhi, S.~Vellucci, M.~Barbuto, A.~Monti, Z.~Hamzavi-Zarghani, L.~Stefanini, D.~Ramaccia, F.~Bilotti, and A.~Toscano, ``Array synthesis of circular huygens metasurfaces for antenna beam-shaping,'' \emph{IEEE Antennas Wirel. Propag. Lett}, 2023.

\bibitem{hashemi2009electronically}
M.~R. Hashemi and T.~Itoh, ``Electronically controlled metamaterial-based leaky-wave transmission-line for conformal surface applications,'' in \emph{2009 IEEE MTT-S International Microwave Symposium Digest}.\hskip 1em plus 0.5em minus 0.4em\relax IEEE, 2009, pp. 69--72.

\bibitem{gomez2011analysis}
J.~L. G{\'o}mez-Tornero, ``Analysis and design of conformal tapered leaky-wave antennas,'' \emph{IEEE Antennas Wirel. Propag. Lett}, vol.~10, pp. 1068--1071, 2011.

\bibitem{lee20232}
H.~Lee and D.-H. Kwon, ``2-d circularly-polarized printed metasurface leaky-wave antennas on a conformal aperture,'' \emph{IEEE Antennas Wirel. Propag. Lett}, 2023.

\bibitem{sleasman2015waveguide}
T.~Sleasman, M.~F. Imani, W.~Xu, J.~Hunt, T.~Driscoll, M.~S. Reynolds, and D.~R. Smith, ``Waveguide-fed tunable metamaterial element for dynamic apertures,'' \emph{IEEE Antennas Wirel. Propag. Lett}, vol.~15, pp. 606--609, 2015.

\bibitem{bayraktar2012slots}
O.~Bayraktar and O.~A. Civi, ``Slots on cylindrical substrate integrated waveguide,'' in \emph{Proceedings of the 2012 IEEE International Symposium on Antennas Propag.}\hskip 1em plus 0.5em minus 0.4em\relax IEEE, 2012, pp. 1--2.

\bibitem{alamzadeh2024computational}
I.~Alamzadeh, T.~Williams, and M.~F. Imani, ``Computational angle of arrival detection using dynamic metasurfaces,'' in \emph{58th Asilomar Conference on Signals, Systems, and Computers}.\hskip 1em plus 0.5em minus 0.4em\relax IEEE, 2024, pp. 602--608.

\end{thebibliography}
}

\vfill

\end{document}